\documentclass[9pt,twocolumn,twoside]{pnas-new}

\templatetype{pnasresearcharticle} 

\usepackage{multirow}
\usepackage{tabularray}
\usepackage{dcolumn}
\usepackage{array}

\newcommand{\ii}{\mathrm{i}}

\usepackage{xcolor}
\definecolor{custom}{RGB}{0,144,226}
\newcommand{\RevisionNov}[1]{{\color{black} #1}}

\begin{document}

\title{Parallel mechanical computing:\\ Metamaterials that can multitask}

\author[1]{Mohamed Mousa}
\author[1,2,*]{Mostafa Nouh}

\affil[1]{Department of Mechanical and Aerospace Engineering, University at Buffalo (SUNY), Buffalo, NY 14260-4400, USA}
\affil[2]{Department of Civil, Structural and Environmental Engineering, University at Buffalo (SUNY), Buffalo, NY 14260-4300, USA}

\leadauthor{Mousa}


\authordeclaration{The authors declare no competing interest.}
\correspondingauthor{\textsuperscript{*}To whom correspondence should be addressed. \\E-mail: mnouh@buffalo.edu}


\begin{abstract}
Decades after being replaced with digital platforms, analogue computing has experienced a surging interest following developments in metamaterials and intricate fabrication techniques. Specifically, wave-based analogue computers which impart spatial transformations on an incident wavefront, commensurate with a desired mathematical operation, have gained traction owing to their ability to directly encode the input in its unprocessed form, bypassing analogue-to-digital conversion. While promising, these systems are inherently limited to single-task configurations. Their inability to concurrently perform multiple tasks, or compute in parallel, represents a major hindrance to advancing conceptual mechanical devices with broader computational capabilities. In here, we present a first attempt to simultaneously process independent computational tasks within the same architected structure. By breaking time invariance in a set of metasurface building blocks, multiple frequency-shifted beams are self-generated which absorb notable energy amounts from the fundamental signal. The onset of these tunable harmonics, enables distinct computational tasks to be assigned to different independent ``channels", effectively allowing an analogue mechanical computer to multitask. 
\end{abstract}

\maketitle
\thispagestyle{empty}
\ifthenelse{\boolean{shortarticle}}{\ifthenelse{\boolean{singlecolumn}}{\abscontentformatted}{\abscontent}}{}

\firstpage[18]{2}



\vspace{0.3cm}

\dropcap{M}echanical computing refers to a form of computing where well-established mechanical mechanisms such as levers or fluidics carry out an intended operation \cite{yasuda2021mechanical}. This form, which relies on both analogue (e.g., sliders, curved surfaces) and discrete mechanisms (e.g., gears, pinwheels), historically predates digital computing, and was used to conduct astronomical calculations \cite{efstathiou2018celestial}, estimate tide heights \cite{fischer1912coast}, and even model the economy \cite{bissell2007historical}, long before digital computers appeared on the scene \cite{tsividis2018not}. Analogue computers in particular are those which exploit a continuous variation in a physical parameter (e.g., voltage or mechanical deformation) to perform a computation \cite{maclennan2007review}. In the modern era, they were effectively replaced by their digital counterparts which addressed the limitations of analogue computers, providing compact, noise-free and high-speed computations. Over the past decade, however, analogue computing research regained impetus owing to recent advances in smart materials and engineered structures (i.e., metamaterials), combined with  additive manufacturing and novel fabrication techniques \cite{zangeneh2021analogue, li2019information}. Today's generation of mechanical computers (MCs) proposes novel ways to process bit abstraction and mechanical logic \cite{Helou2022}, and span a wide portfolio of mechanisms, including rotary joints \cite{al2000dynamics}, conductive polymers \cite{Helou2021}, bistable lattices \cite{ion2017digital}, origami systems \cite{treml2018origami, liu2019invariant}, and various micro-mechanical elements \cite{song2019additively}. 

Tuned wave scattering in dispersive media, especially with the advent of engineered materials and waveguides, provides a rich platform for analogue computing \cite{silva2014performing}. Understandably, the spurt of activity in wave-based computational metamaterials originated in photonic media, owing to ultrafast propagation speeds, allowing analogue photonic computers to outpace digital electronics and operate at subwavelength scales due to small optical wavelengths \cite{wan2021, pors2015, cordaro2019, bao2020, wan2020, zhou2021}. On the other hand, despite the inherently lower speeds of elastoacoustic waves, the value of integrating computational \ \cite{zuo2022} and neuromorphic \cite{moghaddaszadeh2024mechanical} functionality within a self-contained mechanical system has triggered a stream of new paradigms in the domain of wave-based analogue mechanical computers (AMCs), i.e., those which employ targetted scattering of elastic and acoustic waves to carry out a prescribed mathematical operation \cite{lv2021implementing, zangeneh2018performing}. In remote or low access conditions, the ability to sustain a bare minimum of computational readiness is invaluable, especially in applications where both the input and output are encoded and processed in the native physical domain, i.e., elastoacoustic deformations. If the computational input comes from the surrounding environment in the form of direct mechanical stimulation such as vibrational excitations, noise, or impinging pressure waves, or from within the structure itself such as back-scattering from a defect or a local inhomogeneity within the elastic medium, AMCs have the potential to bypass digital-to-analogue conversions which tend to consume significant energy \cite{solli2015analog}. Similarly, AMCs are particularly more transformative when the output itself informs a subsequent task or even computation in the same native language, i.e., mechanically feed or excite the next component of the system through the computed wave profile at the readout plane, thus eliminating the need for output digitization. This need for seamless integration between input/output signals and the computing medium also gives new impetus to mechanical operation within extreme environments where electronic components can be rapidly rendered dysfunctional due to elevated temperatures, ionizing radiation, or high magnetic fields \cite{sauder2017automation, dzhurov2023circuit, moon2019enhancing}.

\begin{figure*}[hbt!]
\centering
\includegraphics[width=17.8cm]{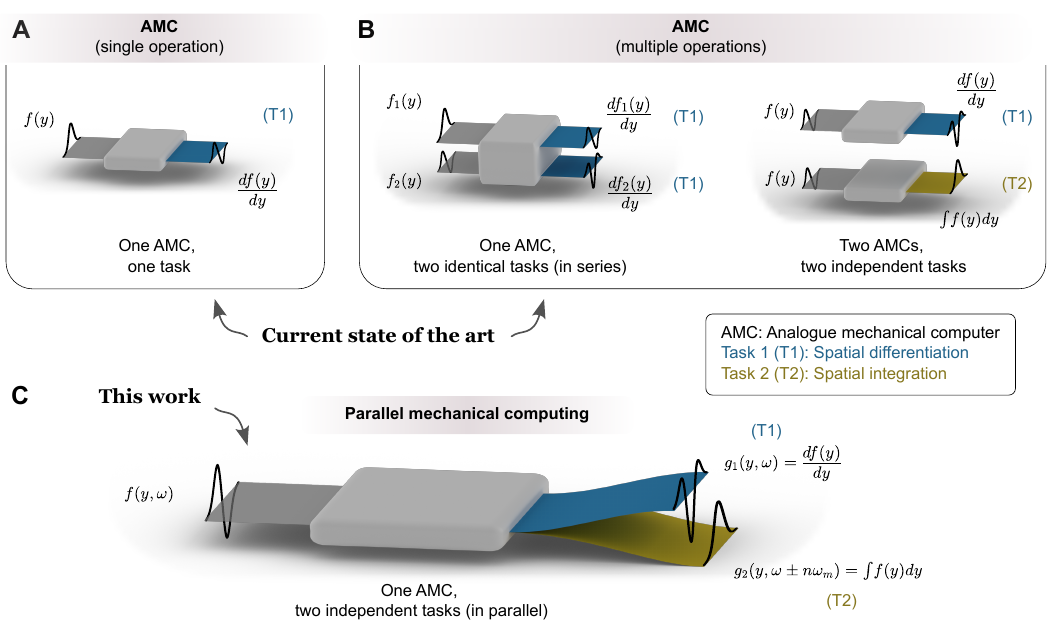}
\caption{\textbf{Wave-based analogue mechanical computing.} (a) Configuration of a basic analogue mechanical computer (AMC). A spatially-encoded input function $f(y)$ morphs into the computed operation (here, a spatial derivative $df(y)/dy$) through wave transformations within an architected structure. (b) In the absence of parallel computing, conducting multiple operations in current AMCs can be either through executing two identical tasks (e.g., spatial differentiation of two different incident signals) one after the other, i.e., in series, or two distinct tasks (e.g., spatial differentiation and integration of an incident signal) via two independent AMCs. (c) System proposed to concurrently run multiple operations on a single input: A monochromatic input function $f(y,\omega)$ at a fundamental frequency $\omega$, undergoes differentiation at the fundamental frequency $\omega$ and integration at a distinct frequency $\omega\pm n\omega_m$, yielding $g_1(y,\omega)$ and $g_2(y,\omega\pm n\omega_m)$, respectively, with both operations conducted in parallel.}
\label{fig1}
\end{figure*}

The emphasis of this work is on a class of AMCs which undertakes complex high-order computations such as differentiation, integration, convolution, and ordinary differential equations by manipulating impinging mechanical waves \cite{zuo2018}. One methodology is the Green’s function metamaterial approach \cite{zhang2019time, dong2018optical}, where the Green’s function of the operator of choice is directly realized in real space using, for instance, phase-shifted Bragg grating \cite{bykov2014optical, liu2017design}, and high-index metamaterials \cite{zangeneh2018performing}. Alternatively, the metasurface (MS) approach \cite{zhou2020analog, chen2017performing, zuo2017mathematical, zuo2018acoustic} adopted here allows the realization of a broader range of complex operators by implementing subwavelength metasurfaces positioned between spatial Fourier transformers, foregoing the need to integrate sophisticated geometries \cite{zangeneh2021analogue}. The aforementioned approaches, while effective, are inherently limited to single-task configurations, as exemplified by Fig.~\ref{fig1}A. As such, to conduct multiple operations, current state-of-the-art AMCs can either execute two identical tasks (e.g., spatial differentiation of two different incident signals) one after the other, i.e., in series, or two distinct tasks (e.g., spatial differentiation and integration of an incident signal) via two independent AMCs, as depicted in Fig.~\ref{fig1}B. This inability to concurrently perform multiple tasks within the same system represents a major obstacle to advancing mechanical computing. Given the intricate configurations of the operator substrates and the number of unit cells involved, the notion of using a new sub-unit for every additional computational task exponentially increases the AMC footprint in a manner that is both inefficient and impractical. As such, crossing this hurdle can be conducive to conceptual devices with broader computational capabilities \cite{rajabalipanah2022parallel}.  

In this work, we present a first successful attempt to concurrently process independent operations within the same architected structure via frequency multiplexing, thus effectively unlocking parallel computing in AMCs. To accomplish this, we will tap into the dynamical features of time-variant periodic media \cite{sapienza2023apl}, building off nonreciprocal wave phenomena and the manipulation of monochromatic incident waves to induce supplementary waves with spectral footprints that are distinct from the primary one \cite{shaltout2019spatiotemporal, wang2020theory}. By exploiting time-modulated metasurfaces as building blocks of an AMC, we will demonstrate the system's ability to instigate multiple scattered wave beams, representing frequency conversions (harmonics) of a single input, that simultaneously propagate within distinct channels along the AMC's operational spectrum. Since these frequencies can be independently tuned as functions of the constitutive parameters of the unit cell geometry and physical properties, they can be potentially assigned distinct computational tasks, thus allowing parallel operations to take place. As a result, upon completion, the computing output can be directly extracted from the culminating frequency-domain waveforms at the AMC's readout plane at the respective frequencies, as graphically shown in Fig.~\ref{fig1}C. The system is based on acoustic metasurface layers structured from the well-established Helmholtz resonator unit cells \cite{fang2006ultrasonic, wang2008acoustic, cheng2008one}. We will show that the configuration of such resonators effectively alters the characteristics of propagating waves, enabling the realization of metasurfaces that can impose the desired transmission and phase profiles and subsequently impart various operations on an incoming wave. 


\section*{Theory}
\subsection*{Metasurface approach} 
\begin{figure*}[hbt!]
\centering
\includegraphics[width=17.8cm]{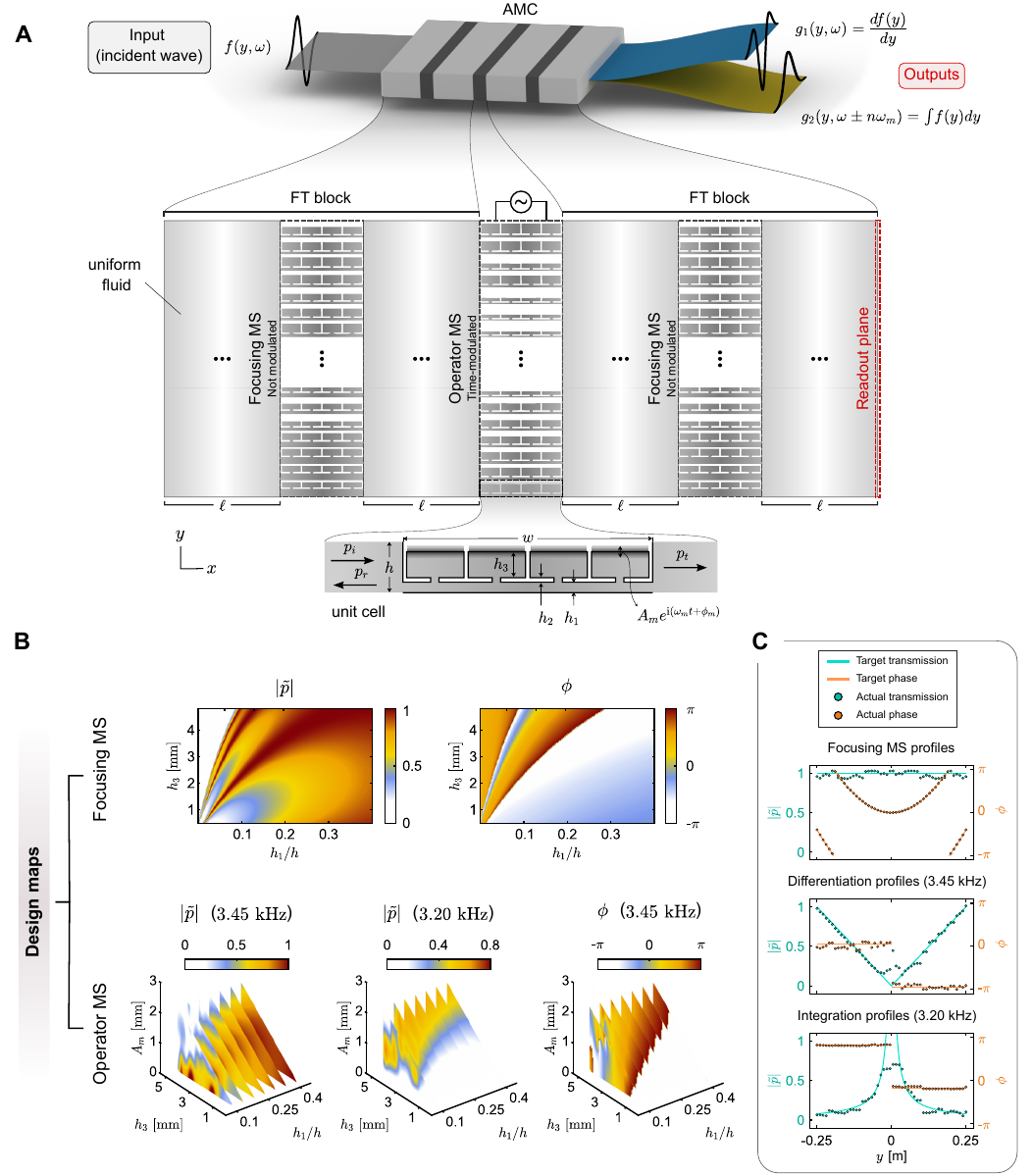}
\caption{\textbf{Transmission and phase profiles at distinct frequency channels.} (a) Schematic diagram of a wave-based analogue mechanical computer (AMC) capable of concurrently executing multiple operations on a single input. The AMC consists of an operator MS sandwiched between two FT blocks, with each metasurface comprising $50$ HR-based unit cells (waveguides). Each unit cell encompasses a straight pipe of length $w$ and height $h_1$, and four resonators of height $h_3$. The top boundaries of the operator MS unit cells are time-modulated with an amplitude $A_m$ and a phase $\phi_m$ at the modulation frequency $\omega_m$. (b) The bottom panel depicts the design maps of the time-modulated operator MS, including the transmission amplitude $|\tilde{p}|$ and phase $\phi$ at a fundamental frequency of $3.45$ kHz and the transmission amplitude of a down-converted frequency of $3.20$ kHz. The top panel depicts the design maps of the focusing MS, which are essentially cross-sections of the fundamental frequency design maps at no modulation, i.e., $A_m=0$. (c) Target and actual profiles of $\tilde{p}$ and $\phi$ for a focusing MS (top), a differentiator MS at the fundamental frequency (middle), and an integrator MS at the down-converted frequency (bottom). It's worth noting that the pressure amplitude $|\tilde{p}|$ is directly proportional to the amplitude of the transmission coefficient $|\bar{T}|$ rather than being exactly equal, which allows for a greater diversity of unit cells satisfying the target design profiles.}
\label{fig2}
\end{figure*}

The analogue mechanical computer (AMC), depicted in Fig.~\ref{fig2}A, consists of three main components: a spatial Fourier transform sub-block (FT), an operator metasurface or a space-filtering sub-block (SF), and an inverse Fourier transform sub-block (IFT), adopting the well-developed metasurface approach \cite{zangeneh2021analogue} wherein a mathematical operation is applied to an input function $f(y)$, spatially encoded in the form of an incident wave, and transforming it to a corresponding output $g(y)$, via the following scheme:
\begin{equation}
\label{eq1}
     g(y)= \textrm{IFT}\big[H(k_y) \cdot \textrm{FT}[f(y)]\big]
\end{equation}
where $k_y$ is the spatial frequency, and $H(k_y)$ is the transfer function between the incident and transmitted fields describing the desired mathematical operation. For instance, the transfer functions associated with differentiation and integration operations are essentially multiplying and dividing the input by $(\ii k_y)$, respectively. In such an AMC, the input wavefield is introduced to the system at a given frequency, henceforth denoted as the fundamental operational frequency. As the wave propagates through the first sub-block, which is in principle a focusing metasurface \cite{goodman2005introduction,zhou2020flat}, it exhibits a Fourier transformation by which the input function gets transferred to the spatial Fourier (frequency) domain. The focusing metasurface makes use of Snell's law and its transmission coefficient $\bar{T}_\mathrm{FT}$ is designed such that:

\begin{equation}
\label{eq2}
     \bar{T}_{\mathrm{FT}}= e^{\ii \frac{2\pi}{\lambda}\sqrt{y^2+\ell_f^2}}
\end{equation}
where $\ii$ is the imaginary unit, $\lambda$ is the wavelength associated with the input wave frequency, and $\ell_f$ is the intended focal length. With $\bar{T}_\mathrm{FT}$ having an absolute value of unity, it can be inferred from the equation that a near-full transmission is required for all $50$ unit cells constituting the focusing metasurface, while satisfying the required phase profile illustrated in Fig.~\ref{fig2}C. The latter is realized by minimizing $|\phi(y)-\phi_\mathrm{FT}(y)|$ via utilizing the available parameters shown in the design maps (Fig.~\ref{fig2}B), with $\phi$ and $\phi_\mathrm{FT}$ being the actual and target phase angles applied at the focusing metasurface unit cells. It is worth noting that the focusing takes effect over the distance from the center of the focusing metasurface to the center of the subsequent sub-block, thus requiring $\ell_f = \ell + w$ for waves to be effectively focused at the intended location, where $\ell$ and $w$ represent the dimensions defined in Fig.~\ref{fig2}A. Although the current design employs a focal length exceeding several wavelengths (see \textit{SI Appendix}, section~S1), the integration of subwavelength focusing techniques offers substantial size reduction capabilities, promoting a compact design \cite{chen2018deep,liu2019subwavelength,ma2022acoustic}.

The FT operation sets up the wave for the mathematical computation which is conducted by the following sub-block, the operator metasurface. By rendering the input function in the frequency domain, calculus-type operations can be carried out via simple algebraic multipliers. This is realized by tailoring the operator metasurface's transmission profile to satisfy the transfer function $H(k_y)$ associated with the operation of interest. For instance, the transmission coefficients required to obtain a first derivative and an integration of the input function are: 
\begin{equation}
\label{eq3}
    \bar{T}_{\mathrm{diff}} = \ii \Big(\frac{2y}{w}\Big)
\end{equation}
and,
\begin{equation}
\label{eq4}
\bar{T}_{\mathrm{int}} = -\ii \Big(\frac{d}{y}\Big)
\end{equation}
\parshape=0
respectively, where $d << w$ is an arbitrary normalizing length \cite{silva2014performing,zuo2017mathematical}. Similar to the focusing layer, the $50$ unit cells of the operator metasurface are carefully selected from the design maps, with the goal of minimizing the deviation of the actual transmission amplitude $|\bar{T}(y)|$ from the target one $|\bar{T}_\mathrm{SF}(y)|$, which corresponds to the operation-specific transmission function, e.g., $\bar{T}_{\mathrm{diff}}$ and $\bar{T}_{\mathrm{int}}$ for spatial differentiation and integration, respectively. We note that while the focusing layer's transmission profile is symmetric, the operator layer's profile is not. This is attributed to the fact that two different phase angles shifted by $\pi$ are required for the metasurface's two halves ($y \in [-\frac{w}{2},0]$ and $y \in [0, \frac{w}{2}]$), as shown in Fig.~\ref{fig2}C. Finally, the wave undergoes another Fourier transformation at the IFT sub-block, which reverts it back to the time domain, and the desired output wave is received at the system's final terminal, i.e., the mechanical computer's readout plane.

\subsection*{Unit cell}
The dynamics governing the AMC necessitate a unit cell with a wide range of both transmission amplitudes $|\bar{T}|$ and phase angles $\phi$. A subwavelength unit cell is adopted for that purpose which consists of a straight pipe coupled with four shunted Helmholtz resonators (HRs), as schematically illustrated in Fig.~\ref{fig2}A \cite{li2015metascreen,li2015three}. To achieve near-full transmission through the unit cell for the focusing metasurfaces (FT and IFT sub-blocks), the straight pipe is constructed as half-wavelength long ($w=\lambda/2$) to match the acoustic impedance of the incident waves. Additionally, the combination of four HRs acting as lumped elements, generates the necessary effective acoustic reactance, enabling the full $2\pi$ phase range to be achieved by tuning the height $h_3$. However, the majority of mathematical operations in the SF sub-block require transmission profiles with varied amplitudes, typically ranging from $0$ to $1$, along with the corresponding full $2\pi$ phase range. By incorporating the height of the straight pipe, $h_1$, as a tunable parameter, we limit the flow partially, allowing for precise control of the transmission amplitude. While the fine tuning of $h_{1,3}$ allows the unit cell to satisfy the aforementioned design requirements, its structural simplicity enables straightforward implementation, facilitating both simulations (\textit{SI Appendix}, Fig.~S2) and experimental realizations \cite{tian2019programmable}. 

Utilizing this unit cell configuration to construct the three AMC components, two separate AMCs are designed to perform the individual mathematical operations of spatial differentiation and integration following the procedure explained earlier. Without loss of generality, we use a Gaussian function, $p_i(y)=ye^{-100y^2}$, as an input function at a frequency of $3.45$ kHz. The latter is equivalent to the resonant frequency of the unit cell's straight pipe, yielding a wavelength of $\approx 0.1~\mathrm{m}$ for atmospheric air. Note that $p_i(y)$ represents shape of the input waveform and that the ensuing process is amplitude-independent. As such, it can be scaled as required for any given application (e.g., amplitude of an incident pressure wave). The results, depicted in \textit{SI Appendix}, Fig.~S3, show the system's ability to accurately execute both operations, and serve as a benchmark for the rest of the study.

\begin{figure*}[hbt!]
\centering
\includegraphics[width=17.8cm]{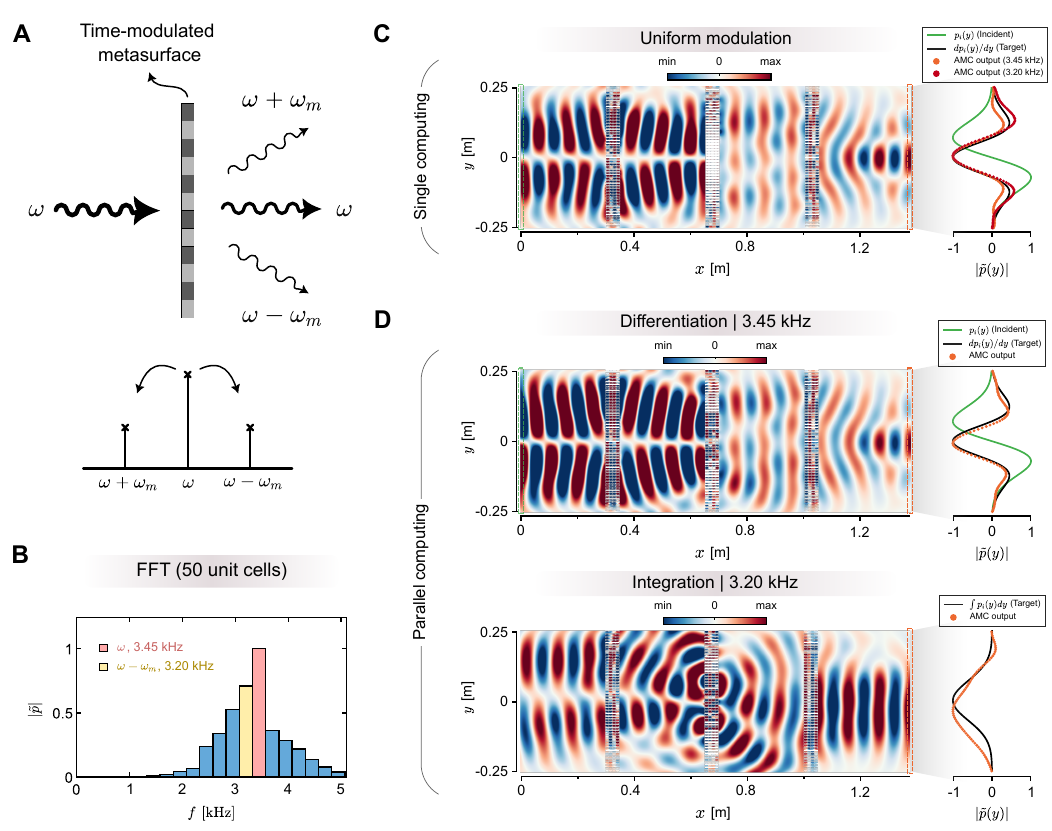}
\caption{\textbf{Parallel mechanical computing.} (a) Conceptual diagram illustrating how the bulk energy of a wave incident on a time-modulated metasurface is redistributed over the fundamental frequency $\omega$ and its harmonics, depending on the modulation frequency $\omega_m$. (b) FFT analyses of the 50 unit cells used by the time-modulated AMC system. Spectral distribution of pressure amplitudes highlights the system's ability to perform parallel operations. The two frequency channels chosen for the two computational tasks are colored differently. (c) Wavefield of an AMC configured to perform differentiation, subject to a uniform modulation. Spatial pressure profile $p_i(y)$ at the AMC's input plane is shown on the right side for reference. The AMC's output at the readout plane at $3.45$ kHz and $3.20$ kHz, as well as the analytically differentiated input (target), are shown, demonstrating the repeated differentiation operation at the fundamental frequency and its down-converted harmonic. (d) Wavefields of the AMC at distinct frequencies confirming successul parallel computing. The top panel displays the wavefield of a differentiator at the fundamental frequency ($3.45$ kHz), while the bottom panel displays the wavefield of an integrator at the down-converted frequency ($3.20$ kHz). Adjacent to each panel, the spatial pressure profiles at the AMC's input and readout plane are shown, along with the analytical output for each operator for comparison.}
\label{fig3}
\end{figure*}

\subsection*{Temporal modulation} The metasurface approach, outlined thus far, is neither capable of carrying out more than one operation at once, nor performing a different operation than the one it is specifically configured for (e.g., differentiation or integration). As such, a single AMC can only undertake two tasks in series (Fig.~\ref{fig1}B), and both tasks have to be identical (e.g., two differentiations or two integrations depending on what the SF block is designed to accomplish). This inability to concurrently perform multiple tasks represents a major hindrance in current state-of-the-art AMCs, further exacerbating the problem of low computational speeds. To unlock parallel computing in AMCs, we exploit targetted changes in the frequency content of a propagating signal associated with periodic media which carry a momentum bias, and stemming from the temporal modulation of its unit cell properties. Across photonic \cite{wang2021space}, acoustic \cite{adlakha2020frequency,kang2022sound,lin2022anomalous}, and elastic phased structures \cite{palermo2020surface,moghaddaszadeh2021nonreciprocal,moghaddaszadeh2022complex}, these modulations have been shown to redistribute notable portions of the bulk wave energy into side bands representing up and down-converted harmonics of the fundamental frequency (Fig.~\ref{fig3}A). To enable these frequency conversions, the height of the shunted resonators in the AMC's unit cell is modulated at a rate, $\omega_m=0.25$ kHz, that is much smaller than the fundamental $3.45$ kHz frequency to ensure a stable response while still allowing wave amplitudes at different frequency channels to be controllable. This dynamic modulation is practically realizable via different methods, e.g., by pumping fluid into and out of the unit cell \cite{tian2019programmable}, without notable complexity. The modulation frequency is kept the same for all unit cells for simplicity.

In our numerical finite element model, each resonator domain is defined as a moving mesh where deformed mesh positions are introduced as degrees of freedom in the system dynamics (\textit{SI Appendix}, section~S4). The top boundaries of the resonators are set to have a reciprocating vertical motion, $A_m \cos{(\omega_m t +\phi_m)}$, where $A_m$ and $\phi_m$ indicate the modulation amplitude and phase, respectively. Side boundaries are set to have zero normal displacement, ensuring the mesh deformation occurs in the desired direction. Parametric studies are then conducted with varying tunable heights $h_1$ and $h_3$, and modulation parameters $A_m$ and $\phi_m$. Specifically, for every configuration, a time-dependent simulation is run with the unit cell embedded within a waveguide structure (\textit{SI Appendix}, Fig.~S5), and a subsequent analysis of the simulation data using a Fast Fourier Transform (FFT) provides discrete values for the transmission amplitudes and phases at select frequencies (See Fig.~\ref{fig2}B).

\section*{Results and Discussion}

\subsection*{Single computing} As a sanity check, and for the sake of future comparison, we first introduce a uniform time modulation to a system that is originally designed to conduct a single mathematical operation, namely differentiation, meaning that the modulation parameters $A_m$ and $\phi_m$ are fixed for all unit cells in the operator MS. Despite the anticipated energy redistribution from the fundamental frequency channel to other harmonics, the AMC's functionality is expected to remain consistent across all frequency channels due to the uniformity of the imposed modulation. A wavefield snapshot of this computer, depicted in Fig.~\ref{fig3}C, illustrates the system's ability to extract the spatial derivative of the input load, i.e., $dp_i(y)/dy$, at the readout plane. Additionally, the rightmost panel of the figure confirms the ability of the AMC to successfully compute $dp_i(y)/dy$ at both the fundamental ($3.45$ kHz) and down-converted ($3.20$ kHz) frequency channels, albeit with a slightly lower amplitude for the latter.

\subsection*{Parallel computing} Following this confirmation, the AMC is now set up to concurrently run two independent operations in parallel, a capability that has been thus far elusive. The modulation parameters as well as the tunable heights are carefully selected such that the highest pressure amplitudes take place at the fundamental ($\omega=3.45$ kHz) and down-converted ($\omega-\omega_m=3.20$ kHz) frequencies, as depicted by the color-coded bars in Fig.~\ref{fig3}B. The goal is for the AMC to fulfil a primary mathematical operation (differentiation) at the fundamental frequency of the input signal, while simultaneously running a secondary operation (integration) encrypted within a ``hidden" layer known to the user, which corresponds to the down-converted frequency. 

The metasurface approach detailed earlier is reproduced, with the FT sub-blocks being identical to those of a single computing system. The SF sub-block is, however, intricately constructed from time-modulated unit cells that satisfy the transmission profiles of two distinct tasks: differentiation and integration at the two select frequencies, respectively. The deliberate selection of these two contrasting operations, as evidenced by their distinct profiles, while imposing substantial challenges on the system's capabilities, establishes the validity of the proposed approach and precludes any potential attribution to randomness. Most importantly, unlike traditional mathematical calculations where solution difficulty scales with problem complexity, the metasurface approach solely depends on the transmission profiles associated with the desired operation. For instance, any transfer function, including the most complex ones, has a transmission amplitude between 0 and $\infty$, and a phase between $-\pi$ and $\pi$. That being so, these two specific operations span the whole range of transmission amplitudes (differentiation $\rightarrow$ 0 to 1, integration $\rightarrow$ 0 to $\infty$) and a broad range of phase shifts, making them very strong candidates for performance evaluation. It's worth noting that the $h_1$, $h_3$, and $A_m$ parameters are simultaneously tuned to control the transmission amplitudes of both channels and the phase of the fundamental channel, while $\phi_m$ is thereafter tuned independently to satisfy the phase profile of the down-converted channel. Since the primary computation is expectedly more prominent in the time-domain wavefield, a high-resolution FFT is applied to the time-dependent study results in the window of simulation time to discern the wavefields at the fundamental, $\omega$, and down-converted, $\omega - \omega_m$, frequencies (\textit{SI Appendix}, section~S6). Figure~\ref{fig3}D shows the extracted outputs at the readout plane at both frequencies, illustrating the AMC's effectiveness in executing the two operations in parallel. The rightmost panels compare the waveforms of the spatial derivative, $dp_i(y)/dy$, and integral, $\smallint p_i(y) dy$, obtained from the AMC with their analytically obtained counterparts. The data shows the system's ability to perform the two intended, and distinct, operations at discrete frequency channels, with pinpoint accuracy for the primary task and a very reasonable accuracy for the secondary one. The entire computation takes the same time as that consumed by the single-task AMC.

\begin{figure*}[hbt!]
\centering
\includegraphics[width=17.8cm]{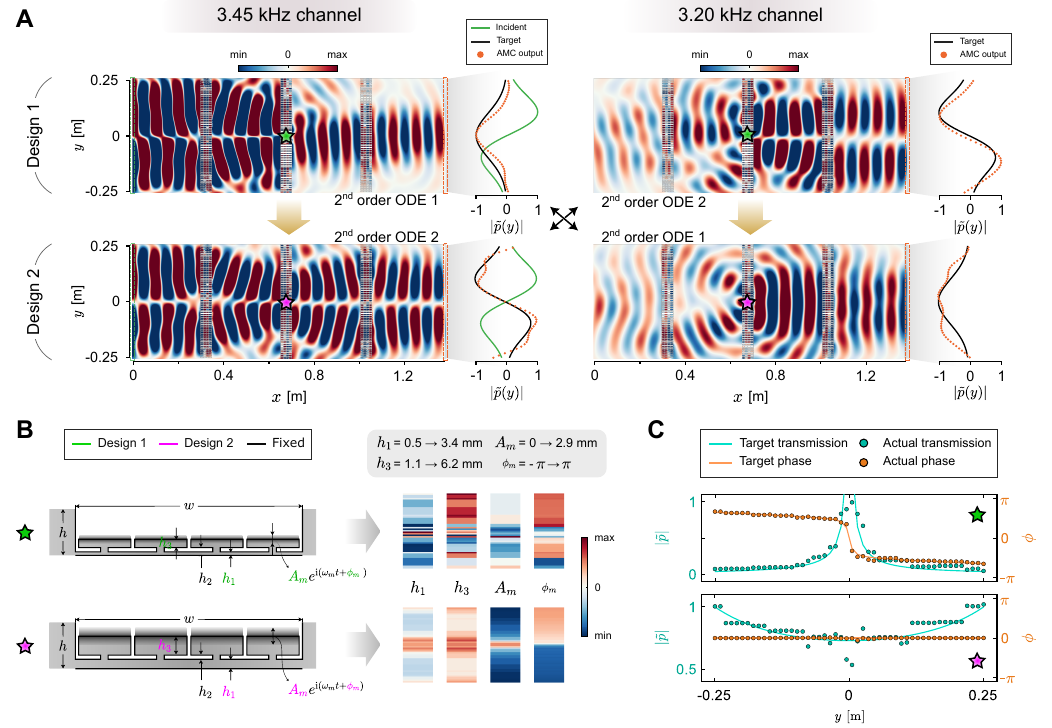}
\caption{\textbf{Robustness and task switchability.} (a) The parallel AMC maintains functionality regardless of the required operation or the frequency channel in which it is carried out. Wavefields in the top row are of an AMC designed to perform two different second-order ODEs of the input waveform at $3.45$ kHz and $3.20$ kHz. The bottom row shows the effectiveness of the AMC when reconfigured to switch the frequencies at which the two operations are implemented. (b) Two sample unit cells and the design parameters, $h_1$, $h_3$, $A_m$, and $\phi_m$, required for the first (top) and second (bottom) designs across each of the 50 unit cells are shown, along with the numerical ranges of each parameter. (c) Target and actual profiles of $\tilde{p}$ and $\phi$ for the first (top) and second (bottom) ODE solvers.}
\label{fig4}
\end{figure*}

\subsection*{Insensitivity to task type} We place emphasis on the fact that this newly-embedded AMC ability to conduct computations in parallel is highly robust. In other words, the system maintains its functionality regardless of the required operation or the frequency channel in which it is implemented. To prove the validity of this claim and better showcase the system's capabilities of tackling complex problems, we present an AMC as a full-fledged ordinary differential equation (ODE) solver. We deliberately select two second-order equations to generate contrasting wavefields for enhanced visualization and comparison, despite the system's capability to handle higher-order ODEs (\textit{SI Appendix}, section~S7), and ask the AMC to solve both of them simultaneously. As shown in the top panel of Fig.~\ref{fig4}A, the reconfigured AMC exhibits excellent performance consistent with that observed during differentiation and integration. To further substantiate the system's robustness and design flexibility, we re-design the AMC, as depicted in the bottom panel, to emulate the aforementioned performance but with the first and second ODEs being performed at the flipped channels, i.e., the down-converted and fundamental frequencies, respectively. Despite substantial variations between the two transmission profiles leading to entirely different designs of the tuned parameters as illustrated in Figs.~\ref{fig4}B~and~\ref{fig4}C, the system maintains its functionality, as depicted in the bottom row of Fig.~\ref{fig4}A. The results confirm the AMC's insensitivity to specific operation profiles and a greater dependence on the available unit cell designs and their inherent performance. In addition to the previously discussed advantages, the FFT of the design unit cells, shown in Fig.~\ref{fig3}B, shows notable transmission amplitudes at several other harmonics along the operational spectrum, indicating the potential to perform more than merely two parallel operations. This capability can be effectively exploited by optimizing the modulation frequency as well as the tuning parameters of the underlying metasurfaces. \RevisionNov{It is however important to note that achieving more than two independent computations would require the modulation of additional unit cell features.}

\RevisionNov{Finally, beyond the assessments of robustness and design flexibility already provided, we discuss in \textit{SI Appendix}, sections~S8~and~S9, the resilience of the proposed methodology in the face of more complex loading conditions that extend beyond a single-frequency signal. Since the AMC receives its input from an incident pressure, the external environment is bound to provide an unfiltered input where the signal of interest is polluted with broadband noise, or even an input signal which inherently contains a broader range of frequencies. In both scenarios, we demonstrate the AMC's continued ability to perform at par with the effectiveness shown here and execute the intended computations in parallel.}

\subsection*{Concluding remarks} In summary, this work has demonstrated the successful realization of parallel computing via frequency multiplexing within the realm of analogue mechanical computers that exploit targetted, guided wave scattering in tuned metasurfaces. The presented system was shown to be capable of concurrently executing two distinct mathematical operations, by breaking off monochromatic incident waves into supplementary signals, two of which were carefully tailored to output the intended computations on the input signal. The process was enabled by the utilization of acoustic metasurface unit cells composed of Helmholtz resonating cavities, the geometrical parameters of which exhibiting finely tuned temporal modulations. Notably, we have shown this approach to be robust and effectively insensitive to the type of computation or the frequency in which it is hosted. It should be noted that while there is added complexity to enable this form of multiplexing, the extra cost, space, and error rate associated with stacking multiple single-task AMCs to perform the same functions exponentially exceeds the complexity brought upon by introducing frequency multiplexing to analogue computers, owing to recent advances in engineered materials and fabrication techniques. In addition to providing a compact solution with minimal trade-offs \RevisionNov{(see \textit{SI Appendix}, section~S10)}, the ability to distinguish between primary computations (which take up the bulk of the energy spectrum and dominate the time domain wavefields; see \textit{SI Appendix}, Movie~S2) and secondary computations (which are hosted in low-observable frequency channels only identifiable to the user) is unique to the proposed AMC and supports a stealth form of computation that is otherwise unattainable through simple replication of conventional sub-units. Finally, the inherent multitasking capability of the shown metamaterial-based system presents exciting avenues for future exploration in the domains of physical and reservoir computing, both in mechanical media and beyond.

\section*{Materials and Methods}

A COMSOL Multiphysics model is employed with the acoustics module to simulate the system's behavior with air as a working fluid. Initially, single waveguides are evaluated to acquire transmission amplitudes and phases across a wide range of unit cell configurations. A moving mesh is incorporated to account for the deformations in the dynamically-modulated resonators. The extracted data is then used to construct design maps. Following the methodology outlined in the theory section, a multitasking AMC is established by identifying optimal unit cell configurations which are chosen to satisfy the prescribed criteria and achieve the desired transmission and phase profiles. Finally, a time-dependent study is conducted to obtain the emergent wavefield, to which a subsequent FFT analysis is applied over the intended frequency range of 0–4 kHz to extract the deformation profiles at the fundamental and secondary operating frequencies, enabling the evaluation of the AMC's parallel computing performance at select frequencies via the multiplexing approach.

\acknow{This work was supported by the US Army Research Office (ARO) under award no. W911NF-23-1-0078 as well as the US National Science Foundation (NSF) under award no. 1847254 (CAREER).}

\showacknow{} 




\vspace{0.25cm}

\bibliography{pnas-sample}

\end{document}